\documentclass[epj,floatfix]{svjour}
\usepackage{graphics}
\begin{document}
\title{Geometric phase of a qubit interacting with a squeezed-thermal bath} 

\author{Subhashish Banerjee\inst{1} \and R. Srikanth \inst{1,2}
}                     
\institute{Raman Research Institute, Sadashiva Nagar,
Bangalore - 560 080, India \and 
Poornaprajna Institute of Scientific Research,
Devanahalli, Bangalore- 562 110, India
}
\date{Received: date / Revised version: date}
%
\abstract{ We study  the geometric phase of an  open two-level quantum
system under the influence of a squeezed, thermal environment for both
non-dissipative    as   well    as    dissipative   system-environment
interactions.  In the non-dissipative case, squeezing is found to have
a similar  influence as  temperature, of suppressing  geometric phase,
while  in the  dissipative  case, squeezing  tends  to counteract  the
suppressive  influence of  temperature in  certain regimes.   Thus, an
interesting feature that emerges from  our work is the contrast in the
interplay between squeezing and thermal effects in non-dissipative and
dissipative  interactions.   This  can  be useful  for  the  practical
implementation  of  geometric   quantum  information  processing.   By
interpreting the open  quantum effects as noisy channels,  we make the
connection  between  geometric   phase  and  quantum  noise  processes
familiar from quantum information theory.
} 
\maketitle

\section{Introduction\label{sec:intro}}

Geometric  Phase  (GP)  brings  about  an  interesting  and  important
connection between phase and the intrinsic curvature of the underlying
Hilbert  space.  In  the   classical  context  it  was  introduced  by
Pancharatnam  \cite{sp56},  who  defined  a phase  characterizing  the
intereference   of    classical   light   in    distinct   states   of
polarization.  Its   quantum  counterpart  was   discovered  by  Berry
\cite{mb84}  for  the  case   of  cyclic  adiabatic  evolution.  Simon
\cite{bs83} showed this to be a  consequence of the holonomy in a line
bundle over parameter space  thus establishing the geometric nature of
the phase.  Generalization of  Berry's work to  non-adibatic evolution
was carried out by Aharonov and Anandan \cite{aa87} and to the case of
non-cyclic  evolution  by  Samuel  and Bhandari  \cite{sb88},  who  by
extending Pancharatnam's ideas for the interference of polarized light
to  quantum mechanics  were able  to make  a comparison  of  the phase
between  any two  non-orthogonal  vectors in  the  Hilbert space.   An
important   development  was   carried  out   by  Mukunda   and  Simon
\cite{ms93}, who, making  use of the fact that GP  is a consequence of
quantum kinematics, and is thus  independent of the detailed nature of
the dynamics in state space, formulated a quantum kinematic version of
GP.

Uhlmann  was  the  first to  extend  GP  to  the case  of  non-unitary
evolution  of mixed  states,  employing the  standard purification  of
mixed  states  \cite{uhlmann}. Sj\"oqvist  {\em  et al.}  \cite{sjo00}
introduced   an   alternate   definition   of  geometric   phase   for
nondegenerate density opertors undergoing unitary evolution, which was
extended by Singh {\em et al.}  \cite{sgh03} to the case of degenerate
density operators.  A kinematic approach  to define GP in mixed states
undergoing nonunitary evolution, generalizing the results of the above
two  works,  has   recently  been  proposed  by  Tong   {\em  et  al.}
\cite{ts04}. Wang et al. \cite{wang06,wang07} defined a GP based on a mapping
connecting density matrices representing  an open quantum system, with
a nonunit vector ray in  complex projective Hilbert space, and applied
it to study the effects of a squeezed-vacuum reservoir on GP.

The geometric nature  of GP provides an inherent  fault tolerance that
makes  it a  useful resource  for  use in  devices such  as a  quantum
computer \cite{dua01}.  There have been proposals to observe GP
in  a Bose-Einstein-Josephson junction \cite{bm05} and in
a superconducting nanostructure \cite{gf00}, and  of using it
to control the evolution of the quantum state \cite{jj00}. However, in
these  situations the effect  of the  environment is  never negligible
\cite{np99}. Also  in the context  of quantum computation,  the qubits
are never isolated but under some environmental influence. Hence it is
imperative  to study GP  in the  context of  Open Quantum  Systems. An
important step  in this  direction was taken  by Whitney {\em  et al}.
\cite{wg03},  who carried  out an  analysis of  the Berry  phase  in a
dissipative   environment   \cite{chi04}.    Rezakhani   and   Zanardi
\cite{rz06} and Lombardo and Villar \cite{lom06} have also carried out
an  open system  analysis of  GP, where  they were  concerned, amongst
other things,  with the interplay  between decoherence and  GP brought
about  by thermal  effects  from the  environment.  Sarandy and  Lidar
\cite{sar06}  have  introduced  a  self-consistent framework  for  the
analysis of Abelian and  non-Abelian geometric phases for open quantum
systems undergoing cyclic adiabatic evolution. The GP acquired by open
bipartite  systems  has recently  been  studied  by  Yi {\em  et  al.}
\cite{yixnjp} using the quantum  trajectory approach. 

In  this  paper we  make  use  of the  method  of  Tong  {\em et  al.}
\cite{ts04} to  study the GP of  a qubit (a  two-level quantum system)
interacting   with  different   kinds  of   system-bath  (environment)
interactions, one  in which  there is no  energy exchange  between the
system  and  its environment,  i.e.,  a  quantum non-demolition  (QND)
interaction    and   one    in   which    dissipation    takes   place
\cite{yipra1,yipra2}.  Throughout,  we assume the  bath to start  in a
squeezed thermal initial state, i.e.,  we deal with a squeezed thermal
bath. The physical  significance of squeezed thermal bath  is that the
decay rate  of quantum coherences in  phase-sensitive (i.e., squeezed)
baths  can be  significantly modified  compared to  the decay  rate in
ordinary  (phase-insensitive)   thermal  baths  \cite{kw88,kim93,bg06}.   A
method to  generate GP  by making  use of a  squeezed vacuum  bath has
recently been proposed by Carollo {\em et al.}  \cite{cor06}.

The open system effects studied  below can be given an operator-sum or
Kraus   representation  \cite{kraus}.    In  this   representation,  a
superoperator ${\cal  E}$ due to environmental  interaction, acting on
the state of the system is given by
\begin{equation}
\label{eq:kraus}
\rho \longrightarrow {\cal E}(\rho) = 
\sum_k \langle e_k|U(\rho \otimes |f_0\rangle\langle f_0|)U^{\dag}|e_k\rangle
= \sum_j E_j \rho E_j^{\dag},
\end{equation}
where $U$ is  the unitary operator representing the  free evolution of
the system,  reservoir, as  well as the  interaction between  the two,
$\{|f_0\rangle\}$   is   the    environment's   initial   state,   and
$\{|e_k\rangle\}$ is a basis  for the environment. The environment and
the system  are assumed to  start in a  separable state. In  the above
equation,  $E_j   \equiv  \langle  e_k|U|f_0\rangle$   are  the  Kraus
operators,   which   satisfy   the  completeness   condition   $\sum_j
E_j^{\dag}E_j = \mathcal{I}$.  The  operator sum representation is not
unique.   Every (infinitely  many)  possible choice  of tracing  basis
$\{|e_k\rangle\}$  in Eq.   (\ref{eq:kraus}) yields  a  different, but
equivalent and  unitarily related, set  of Kraus operators. It  can be
shown  that  any   transformation  that  can  be  cast   in  the  form
(\ref{eq:kraus}) is a completely positive (CP) map \cite{nc00}.

From the viewpoint of quantum communication, these open quantum system
effects correspond  to noisy quantum  channels, and are recast  in the
Kraus representation. We find that  some of them may be interpreted in
terms  of  familiar noisy  quantum  channels.   This abstraction  will
enable us  to connect noisy channels  directly to their  effect on GP,
bypassing  system-specific details.  Visualizing  the effect  of these
channels on GP in a Bloch  vector picture of these open system effects
helps to interpret our GP results in a simple fashion.

The structure of the paper  is as follows. In Section \ref{sec:recap},
we briefly discuss QND open  quantum systems and collect some formulas
which would be of use  later. In Section \ref{sec:gpqnd}, we study the
GP of  a two-level system in  QND interaction with its  bath.  Here we
consider two different kinds of baths.  In Section \ref{sec:bathhc}, a
bath   of  harmonic   oscillators   is considered, and we also briefly
touch upon a  bath of two-level systems.  In Section
\ref{sec:qndkraus}, we point out that  the GP results obtained in this
section  are generic  for any  purely dephasing  channel.   In Section
\ref{sec:nonqnd},  we  study  the  GP  of  a  two-level  system  in  a
dissipative  bath.   Section   \ref{sec:gpbma}  considers  the  system
interacting  with   a  bath  of  harmonic  oscillators   in  the  weak
Born-Markov,   rotating-wave    approximation   (RWA).    In   Section
\ref{sec:bma},  we point  out that  the  GP results  obtained in  this
section  are generic  for any  squeezed generalized  amplitude damping
channel  \cite{srisub}, of  which the  familiar  generalized amplitude
damping  channel   \cite{nc00}  is  a  special  case.    We  make  our
conclusions in Section \ref{sec:discu}.

\section{QND open quantum systems - A recapitulation
\label{sec:recap}} 

To  illustrate the  concept of  QND open  quantum systems  we  use the
percept   of  a   system   interacting  with   a   bath  of   harmonic
oscillators.  Such a  model, for  a two-level  atom, has  been studied
\cite{unr95,pal96,div95} in the context of influence of decoherence in
quantum computation.  We will consider the following Hamiltonian which
models the interaction of a system with its environment, modelled as a
bath of harmonic oscillators, via a QND type of coupling \cite{bg06}
\begin{eqnarray}
H &=& H_S + H_R + H_{SR} \nonumber \\ &=& H_S + 
\sum\limits_k \hbar \omega_k b^{\dagger}_k b_k + H_S 
\sum\limits_k g_k (b_k+b^{\dagger}_k) \nonumber \\
&+& H^2_S \sum\limits_k {g^2_k \over \hbar \omega_k}. 
\label{2a} 
\end{eqnarray}
Here  $H_S$, $H_R$  and $H_{SR}$  stand  for the  Hamiltonians of  the
system   ($S$),  reservoir   ($R$)   and  system-reservoir   ($S$-$R$)
interaction,  respectively. The last  term on  the right-hand  side of
Eq.  (1) is a  renormalization inducing  `counter term'.  Since $[H_S,
H_{SR}]=0$,  (1)  is of  QND  type. Here  $H_S$  is  a generic  system
Hamiltonian  which we  will use  in the  subsequent sections  to model
different physical  situations. The  system plus reservoir  complex is
closed obeying a unitary evolution given by
\begin{equation}
\rho (t) = e^{-{i \over \hbar}Ht} \rho (0) e^{{i \over 
\hbar}Ht} , \label{2b} 
\end{equation}
where
$\rho (0) = \rho^s (0) \rho_R (0)$,
i.e., we assume separable initial conditions. Here we assume 
the reservoir to be initially in a squeezed thermal state, i.e., 
a squeezed thermal bath, with an initial density matrix $\rho_R 
(0)$ given by 
\begin{equation}
\rho_R(0) = S(r,\Phi) \rho_{\rm th} 
S^{\dagger} (r,\Phi),\label{2d} 
\end{equation}
where
$\rho_{\rm th} = \prod_k \left[ 1 - e^{-\beta \hbar \omega_k} 
\right]\exp\left(-\beta \hbar\omega_k b^{\dag}_kb_k\right)$
is the density matrix of the thermal bath, and 
$$S(r_k, \Phi_k) = \exp \left[ r_k \left( {b^2_k 
\over 2} e^{-i2\Phi_k} - {b^{\dagger 2}_k \over 2} 
e^{i2\Phi_k} \right) \right]$$
is the squeezing operator with $r_k$, $\Phi_k$ being the 
squeezing parameters \cite{cs85}. In an open system analysis we 
are interested in the reduced dynamics of the system of 
interest $S$ which is obtained by tracing over the bath degrees 
of freedom. Using Eqs. (\ref{2a}) and (\ref{2b}) 
and tracing over the bath we obtain the reduced density matrix 
for $S$, in the system eigenbasis, as \cite{bg06} 
\begin{equation}
\rho^s_{nm} (t)  =  e^{-{i \over \hbar}(E_n-E_m)t} e^{
i(E^2_n-E^2_m) \eta(t)}
e^{- (E_n-E_m)^2 \gamma(t)} 
\rho^s_{nm} (0). \label{2g} 
\end{equation}
Here
\begin{equation}
\eta (t) = - \sum\limits_k {g^2_k \over \hbar^2 \omega^2_k} 
\sin (\omega_k t), \label{2i} 
\end{equation}
and
\begin{eqnarray}
\gamma (t) &=& {1 \over 2} \sum\limits_k {g^2_k \over \hbar^2 
\omega^2_k} \coth \left( {\beta \hbar \omega_k \over 2} \right) 
\left| (e^{i\omega_k t} - 1) \cosh (r_k) \right. \nonumber \\
&+& \left. (e^{-i\omega_k t} - 
1) \sinh (r_k) e^{i2\Phi_k} \right|^2. \label{2j} 
\end{eqnarray}
For the case of an Ohmic bath with spectral density
$I(\omega) = {\gamma_0 \over \pi} \omega e^{-\omega/\omega_c}$, 
where $\gamma_0$ and $\omega_c$ are two bath parameters, $\eta 
(t)$ and $\gamma (t)$ have been evaluated in \cite{bg06}, where
we have for simplicity taken the squeezed bath parameters 
as
\begin{eqnarray}
\cosh \left( 2r(\omega) \right) &=& \cosh (2r),
\sinh \left( 2r (\omega) \right) = \sinh (2r),\nonumber \\
\Phi(\omega) &=& a\omega, \nonumber
\end{eqnarray}
with $a$ being a constant depending upon the squeezed bath. We 
will make use of Eqs. (\ref{2i}) and (\ref{2j})
in the subsequent analysis (cf. Ref. \cite{bg06} for details). Note that 
the results pertaining to a thermal bath can be obtained from 
the above equations by setting the squeezing parameters $r$ and 
$\Phi$ (i.e., $a$) to zero. 

\section{GP of two-level system in QND interaction with bath
\label{sec:gpqnd}}

In this section we study the GP of a two-level system in QND interaction
with its environment (bath). We consider two classes of baths, one being
the commonly used bath of harmonic oscillators \cite{lom06}, and
the other being a localized bath of two-level systems.

\subsection{Bath of harmonic oscillators \label{sec:bathhc}}

The total Hamiltonian of the $S + R$ complex has the same form as in
Eq. (\ref{2a}) with the system Hamiltonian 
$H_S = {\hbar \omega \over 2} \sigma_3$,
where $\sigma_3$ is the usual  Pauli matrix.  We will be interested in
obtaining  the  reduced dynamics  of  the  system.   This is  done  by
studying the reduced  density matrix of the system  whose structure in
the  system  eigenbasis is  as  in  Eq.  (\ref{2g}).  For  the  system
described by $H_S$ an appropriate  eigenbasis  is given  by the  Wigner-Dicke
states   \cite{rd54,jr71,at72}  $|j,   m  \rangle$,   which   are  the
simultaneous eigenstates  of the angular momentum  operators $J^2$ and
$J_Z$, and we have
$H_S|j, m \rangle  = \hbar \omega  m |j, m \rangle 
=  E_{j,m} |j, m \rangle$.
Here $-j \leq m \leq j$. 
For the two-level system considered here, $j = {1 \over 2}$ and 
hence $m = -{1 \over 2}, {1 \over 2}$.
Using this basis in Eq. (\ref{2g}) we obtain the reduced density matrix 
of the system as 
\begin{eqnarray}
\rho^s_{jm,jn}(t) &=& e^{-i \omega (m-n)t} e^{i(\hbar 
\omega)^2 (m^2 - n^2) \eta(t)} \nonumber \\ &\times &
e^{-(\hbar \omega)^2 (m - n)^2 \gamma(t)} 
\rho^s_{jm,jn}(0). \label{3c} 
\end{eqnarray} 
It follows from Eq. (\ref{3c})  that the diagonal  elements of the
reduced density matrix signifying  the population remain unaffected by
the  environment whereas the  off-diagonal elements  decay. This  is a
feature  of  the  QND   nature  of  the  system-environment  coupling.
Initially we choose the system to be in the state
\begin{equation}
|\psi(0)\rangle = \cos({\theta_0 \over 2}) |1\rangle +
e^{i \phi_0}
\sin\left({\theta_0 \over 2}\right) |0\rangle. \label{3e}
\end{equation}
Using this we can write Eq. (\ref{3c}) as 
\begin{eqnarray}
\rho^s_{j0,j0}(t) &=& \cos^2({\theta_0 \over 2}) \nonumber \\
\rho^s_{j0,j1}(t) &=& {1 \over 2} \sin(\theta_0) 
e^{-i (\omega t + \phi_0)} e^{-(\hbar \omega)^2 \gamma(t)}
\nonumber \\
\rho^s_{j1,j0}(t) &=&   {1 \over 2} \sin(\theta_0)   
e^{i(\omega t + \phi_0)} e^{-(\hbar \omega)^2 \gamma(t)} \nonumber \\
\rho^s_{j1,j1}(t) &=&  \sin^2({\theta_0 \over 2})
\label{3f}
\end{eqnarray}
We will make use of Eq. (\ref{3f}) to obtain the GP of the above open
system using the prescription of Tong {\em et al.} \cite{ts04}
\begin{eqnarray} 
\Phi_{\rm GP} &=& \arg\left( \sum\limits_{k= 1}^{N} \sqrt{\lambda_k (0)
\lambda_k (\tau)} \langle \Psi_k (0)| \Psi_k (\tau) \rangle
\times \right. \nonumber \\
& & \left.
e^{- \int_0^{\tau} dt \langle \Psi_k (t)| \dot{\Psi}_k (t) \rangle}
\right). \label{3g}
\end{eqnarray}
Hereafter we will consider for GP a quasi-cyclic path where time ($t$)
varies from 0 to $\tau = 2\pi/\omega$, $\omega$ being the system 
frequency.
In the above equation the overhead
dot refers to derivative with respect to time 
and $\lambda_k (\tau)$, $\Psi_k (\tau)$ refer to the eigenvalues
and the corresponding eigenvectors, respectively, of the reduced
density matrix given here by Eq. (\ref{3f}). The eigenvalues of
Eq. (\ref{3f}) are
\begin{equation}
\lambda_{\pm} (t) = {1 \over 2} \left[ 1 + \cos(\theta_0) \epsilon_{\pm} (t)
\right], \label{3h}
\end{equation}
where
$\epsilon_{\pm} (t) = \pm \sqrt{1 + \tan^2(\theta_0)  
e^{-2(\hbar \omega)^2 \gamma(t)}}$.
Since $\gamma(t) = 0$ for $t = 0$, we can see from the above equations
that $\lambda_+ (0) = 1$ and $\lambda_- (0) = 0$. From the structure of 
the Eq. (\ref{3g}) we see that only the eigenvalue $\lambda_+$ and its 
corresponding eigenvector $|\Psi_+ \rangle$ need be considered for the
GP. This normalized eigenvector is found to be 
\begin{equation}
|\Psi_+ (t) \rangle = 
\sin\left({\theta_t \over 2}\right)
|1\rangle + e^{i (\omega t + \phi_0)}\cos\left({\theta_t \over 2}\right)
|0\rangle, \label{3j}
\end{equation}
where 
$\sin\left(\theta_t/2\right) = 
\sqrt{\epsilon_+ + 1 \over 2 \epsilon_+}$.
It can be seen that for $t = 0$, $\sin\left(\frac{\theta_t}{2}\right) 
\rightarrow
\cos\left(\frac{\theta_0}{2}\right)$ 
and $\cos\left(\frac{\theta_t}{2}\right) \rightarrow
\sin\left(\frac{\theta_0}{2}\right)$, as expected. 
Now we make use of Eqs. (\ref{3h}),
(\ref{3j}) in Eq. (\ref{3g}) to obtain GP as 
\begin{eqnarray}
\Phi_{\rm GP} &=& \arg\Big[\left\{{1 \over 2} \left( 1 + \cos(\theta_0) 
\sqrt{1 + \tan^2(\theta_0)  
e^{-2(\hbar \omega )^2 \gamma(\tau)}}\right)\right\}^{1 \over 2} \nonumber\\
&\times&\left\{\cos({\theta_0 \over 2})
\sin\left({\theta_{\tau} \over 2}\right) +  
e^{i \omega \tau} \sin\left(\frac{\theta_0}{2}\right)
\cos({\theta_{\tau} \over 2})\right\} \nonumber \\
&\times& e^{-i \omega  \int_0^{\tau} dt \cos^2({\theta_t \over 2})}
\Big]. \label{3m}
\end{eqnarray}
Here $\gamma(t)$ is as given in
Ref. (\cite{bg06}) for a zero temperature ($T$) bath 
or high $T$ bath.
It can be easily seen from Eq. (\ref{3m}) that if we set the influence
of the environment, encapsulated here by the expression $\gamma(t)$, 
to zero, we obtain for $\tau = {2 \pi \over \omega}$, 
$\Phi_{GP} = -\Omega/2 = -\pi(1- \cos(\theta_0))$,
where $\Omega$ is solid angle subtended by the tip
of the Bloch vector on the Bloch sphere, which is the standard result 
for the unitary evolution of an intial pure state.
More generally, unitary evolution of mixed states also has a simple
relation to the solid angle, given by 
\begin{equation}
\Phi_{\rm GP} = -\tan^{-1}
\left(L \tan\frac{\Omega}{2}\right),
\label{eq:mixyun}
\end{equation} 
where $L$ is the length of the Bloch vector \cite{sjo00,sgh03}.

\begin{figure}
\resizebox{0.95\columnwidth}{!}{\includegraphics{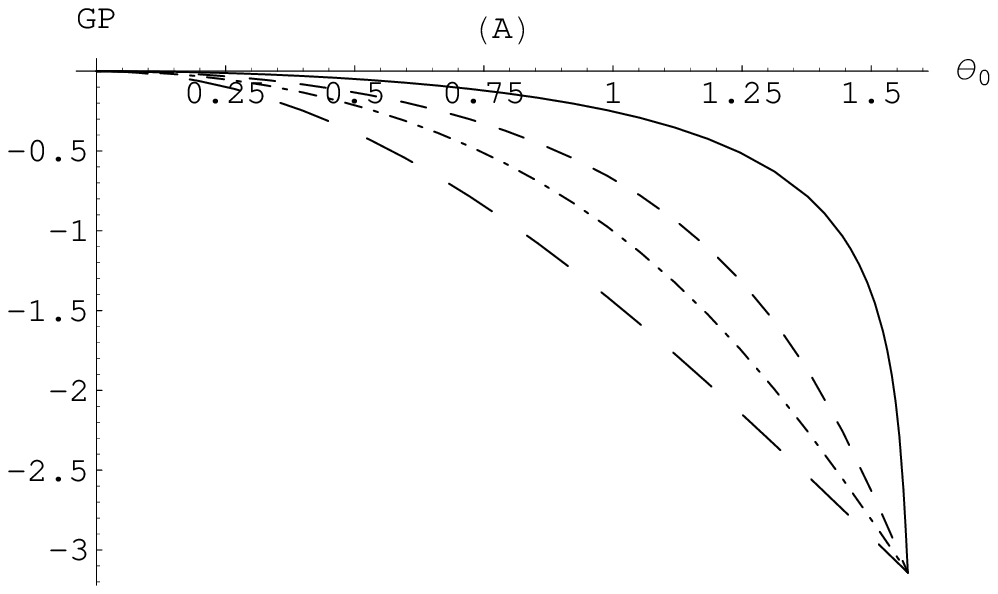}}
\resizebox{0.95\columnwidth}{!}{\includegraphics{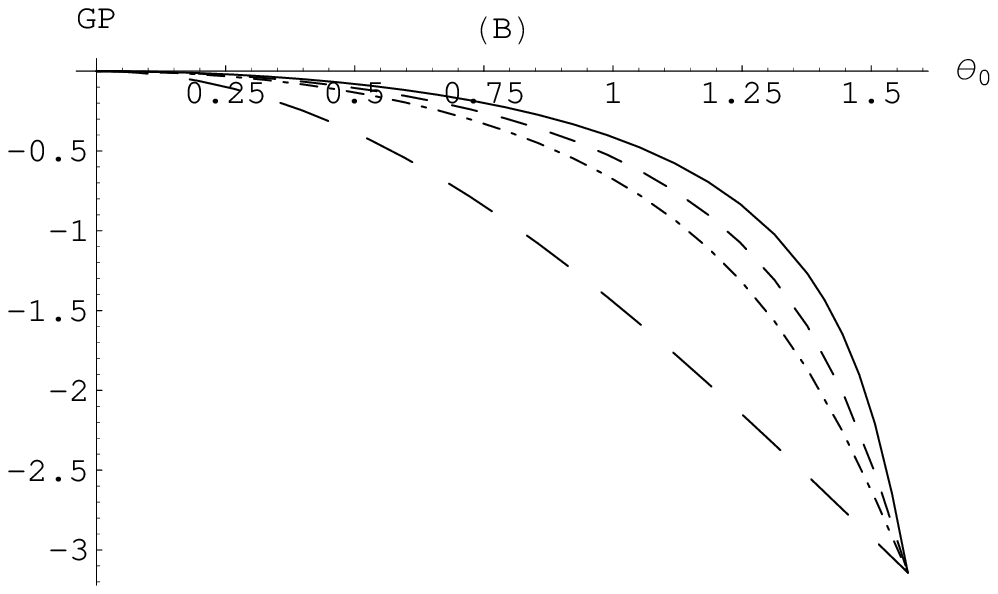}}
\caption{GP (Eq. (\ref{3m}))
as a function of $\theta_0$ (in radians) for different 
temperatures and squeezing at $\gamma_0 =0.0025$. 
In both plots, unitary evolution is depicted by the large-dashed
curve. (A) GP at $r=a=0.0$; 
the dot-dashed, small-dashed and solid curves correspond,
respectively, to temperatures 50, 100, 300.
(B) GP at $T=100$ and $a=0$;
the dot-dashed, small-dashed and solid curves correspond,
respectively, to squeezing parameter $r =$ 0, 0.4, 0.6.
For QND interactions, in the region $\pi/2 < \theta_0 
\le \pi$, the pattern is symmetric but sign reversed. Observe that,
as is true for all QND cases, GP vanishes at $\theta_0 = 0$. This can
be attributed to the fact that the qubit's evolution sweeps no solid angle
in this case. Here, as in all other Figures, we take $\omega = 1$, and
for all Figures in this Section, $\omega_c = 40\omega$.}
\label{fig:qndHO_T0_gam1370}
\end{figure}

\begin{figure}
\resizebox{0.95\columnwidth}{!}{\includegraphics{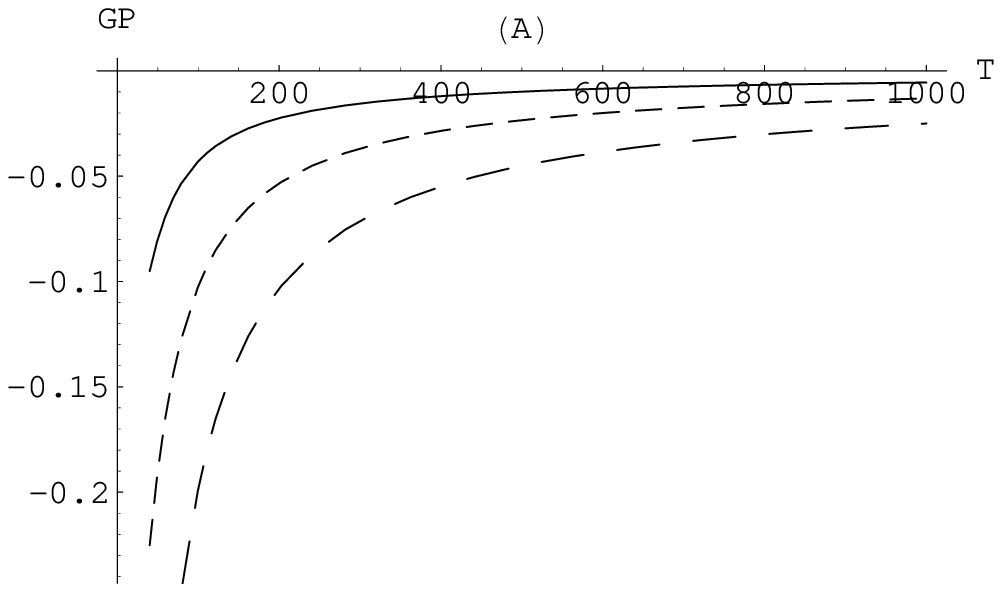}}
\hfill
\resizebox{0.95\columnwidth}{!}{\includegraphics{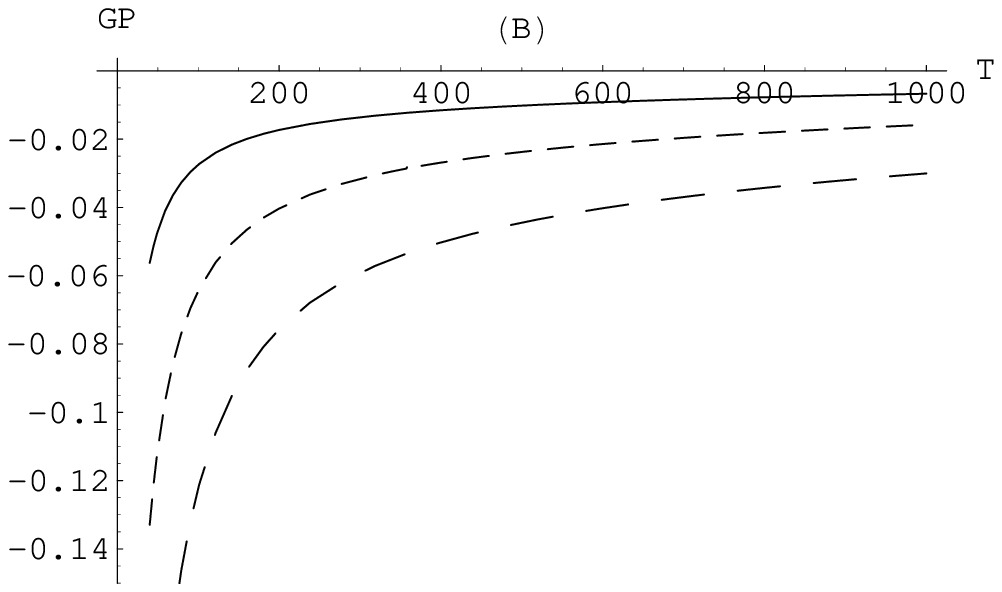}}
\caption{GP (in radians) as a function of temperature ($T$, in units where
$\hbar \equiv k_B \equiv 1$) for QND
interaction with a bath of harmonic oscillators (Eq. (\ref{3m})).
(A) with $\gamma_0 = 0.005$ and vanishing squeezing. 
The solid, dashed and larger-dashed lines
correspond to $\theta_0 = \pi/8, 3\pi/16$ and $\pi/4$.
(B) Same as Figure (A), except that here squeezing is non-vanishing, with
$r=0.7$ and $a=0.1$.}
\label{fig:zantemp}
\end{figure}

The effect of temperature and squeezing on GP is brought out
by Figs. \ref{fig:qndHO_T0_gam1370} and \ref{fig:zantemp}.
From Figs. \ref{fig:qndHO_T0_gam1370}(A) and (B), we see, respectively, that
increasing the  temperature and squeezing induce a departure from
unitary behavior by suppressing GP, 
except at polar angles $\theta_0 = 0, \pi/2$ of
the Bloch sphere. It can be shown that, similarly, 
increase in the $S$-$R$ coupling strength, 
modelled by $\gamma_0$, also tends to suppress GP. 
(Throughout this article,
the Figures use $\omega = 1$. Further, Figures in this Section
use $\omega_c = 40\omega$.) 
The suppresive influence of temperature on GP is also
seen in Figs. \ref{fig:zantemp}, where temperature is varied for
fixed $\theta_0$ and squeezing. A similar 
suppresive influence of squeezing on GP
is brought out by comparing Figs. \ref{fig:zantemp}(A) and 
\ref{fig:zantemp}(B).
These observations are easily interpreted in the Bloch
vector picture, as we discuss later in this section.

Another interesting case is that of qubit subjected to
a bath of two-level systems, studied 
by Shao and collaborators in the context of QND systems \cite{sgc96},
and quantum computation \cite{sh98}. It has also been used
to model a nanomagnet coupled to nuclear and paramagnetic spins
\cite{ps00}. It can be shown \cite{srigp} that this case 
is mathematically similar to that of QND
interaction with a vacuum 
bath of harmonic oscillators for weak $S$-$R$ coupling, 
and hence the dependence of GP on 
$\theta_0$ and $\gamma_0$ is similar to the analogous case discussed
above.

\subsection{Evolution of GP in a phase damping channel\label{sec:qndkraus}}

While the results derived above are for QND $S$-$R$ interactions
with two types of baths, they are quite general, and
in fact apply to any open system effect that can be characterized
as a phase damping channel \cite{nc00}. 
This is a
uniquely non-classical  quantum mechanical noise process, describing
the loss  of quantum  information without the  loss of energy.
This system can be represented by the Kraus operator elements
\begin{equation}
\label{eq:qndkraus}
E_0 \equiv \left[\begin{array}{cc} 1 & 0 \\ 0 & 
e^{i\beta(t)}\sqrt{1-\lambda(t)} 
\end{array}\right], \hspace{1.0cm}
E_1 \equiv \left[\begin{array}{cc} 0 & 0 \\ 0 & \sqrt{\lambda(t)} 
\end{array}\right],
\end{equation}
where  $\beta(t)$  encodes  the  free  evolution  of  the  system  and
$\lambda(t)$  the effect  of  the  environment. 
It is not difficult to see that the QND interactions we
have considered realize a phase damping channel. 

In the case of  QND interaction with a  bath of harmonic
oscillators  (Sec. \ref{sec:bathhc}), 
it is straightforward to verify that with the identification
\begin{equation}
\label{eq:qnd_ho_lambda}
\lambda(t) = 1 - \exp\left[- 2(\hbar\omega)^2 \gamma(t)\right];
\hspace{0.5cm} \beta(t) = \omega t.
\end{equation}
the operators (\ref{eq:qndkraus}) acting on the state
(\ref{3e}) reproduce the evolution
Eq. (\ref{3f}) by means of the map Eq. (\ref{eq:kraus}).
Similarly, the effect of QND interaction with a bath of two level
systems can also be represented as phase damping channel \cite{srigp}.
Our result is in agreement with that of Ref. \cite{wang06}, where
GP is shown to depend on the dephasing parameter, introduced
phenomenologically. Our result is obtained
from a microscopic model, governed by Eqs. (\ref{2a})--(\ref{2d}),
that takes into consideration the interaction of a qubit with a
squeezed thermal bath, the resulting dynamics being shown above to be
equivalent to a phase damping channel. 

In the case of QND interaction, 
any initial state not located on the
$\sigma_3$-axis tends to inspiral towards it, its trajectory remaining
coplanar on the  $x$-$y$ plane. Consequently,  
the entire Bloch  sphere shrinks
into  a prolate  spheroid,  with its  axis  of symmetry  given by  the
$\sigma_3$ axis.  
The   extent  of  inspiral  depends   upon  the  parameter
$\lambda(t)$; the  greater is $\lambda(t)$, the more  is the inspiral.
Greater  squeezing and higher  temperature accentuate  this shrinking.

Guided  qualitatively by  the relation  Eq.  (\ref{eq:mixyun})  we may
interpret GP as directly dependent  on the Bloch vector length $L(t)$,
and the  solid angle ($\Omega$) subtended  at the center  of the Bloch
sphere during a cycle  in parameter space.  Increasing $T$, $\gamma_0$
or  squeezing  results  in  a  larger degree  of  inspiral  causing  a
reduction of both  $L$ and $\Omega$, and hence  greater suppression of
GP relative to the case of unitary evolution.

In Figs. \ref{fig:qndHO_T0_gam1370}(A) and (B), we noted that 
the GP remains invariant at polar angles
$\theta_0 = 0$ and $\theta_0 = \pi/2$. In the case $\theta_0=0$, 
the  Bloch vector remains a constant
$(0,0,1)$ throughout the evolution and hence accumulates no GP. In the
case $\theta_0=\pi/2$, 
note that $\Omega = 2\pi$. From Eq. (\ref{eq:mixyun}), we
see that  irrespective of  the length of  the Bloch vector,  GP should
remain  the same, i.e., $-\pi$.  This suggests  that in  the general
nonunitary  case, when  the  Bloch vector  rotates  on the  equitorial
plane, GP is unaffected by whether  or not there is an inspiral of the
Bloch vector.

The    fall    of    GP     as    a    function    of    $T$    (Figs.
\ref{fig:qndHO_T0_gam1370}(A) and \ref{fig:zantemp}) can be attributed
to  the  fact that  as  $T$  increases the  tip  of  the Bloch  vector
inspirals more  rapidly towards the  $\sigma_3$ axis, and  thus sweeps
less GP.  Squeezing has the same effect as temperature, of contracting
the  Bloch  sphere  along  the  $\sigma_3$ axis,  leading  to  further
suppression   of    GP   (Figs.    \ref{fig:qndHO_T0_gam1370}(B)   and
\ref{fig:zantemp}(B)).

\section{GP of two-level system in non-QND interaction with bath
\label{sec:nonqnd}}

In this  section we study  the GP of  a two-level system in  a non-QND
interaction with  its bath which we  take as one  composed of harmonic
oscillators. We  consider the  case of the  system interacting  with a
bath  which is  initially in  a squeezed  thermal state,  in  the weak
coupling Born-Markov RWA.

\subsection{System interacting with bath in the weak Born-Markov 
RWA \label{sec:gpbma}}

Now  we take  up the  case of  a two-level  system interacting  with a
squeezed  thermal   bath  in  the  weak   Born-Markov,  rotating  wave
approximation. This kind  of system-reservoir ($S{-}R$) interaction is
consonant with the realization that in order to be able to observe GP,
one  should  be in  a  regime  where  decoherence is  not  predominant
\cite{wg03,rz06}. The  system Hamiltonian is $H_S$ and
it  interacts with  the bath  of harmonic  oscillators via  the atomic
dipole operator which in the interaction picture is given as
\begin{equation} 
\vec{D}(t) = \vec{d} \sigma_- e^{-i\omega t} + \vec{d^*}
\sigma_+ e^{i\omega t}, \label{4a} 
\end{equation}
where  $\vec{d}$  is the  transition  matrix  elements  of the  dipole
operator.  The evolution of the reduced density matrix operator of the
system  $S$  in  the   interaction  picture  has  the  following  form
\cite{sz97,bp02}
\begin{eqnarray}
&& \frac{d}{dt}\rho^s(t) = \gamma_0 (N + 1) \nonumber \\
&\times & \left(\sigma_-  \rho^s(t)
\sigma_+ - {1 \over 2}\sigma_+ \sigma_- \rho^s(t) -
{1 \over 2} \rho^s(t) \sigma_+ \sigma_- \right) \nonumber\\
& + & \gamma_0 N \left( \sigma_+  \rho^s(t)
\sigma_- - {1 \over 2}\sigma_- \sigma_+ \rho^s(t) -
{1 \over 2} \rho^s(t) \sigma_- \sigma_+ \right) \nonumber\\
& - & \gamma_0 M   \sigma_+  \rho^s(t) \sigma_+ -
\gamma_0 M^* \sigma_-  \rho^s(t) \sigma_-. \label{4b} 
\end{eqnarray}
Here $\gamma_0$ is the spontaneous emission rate given by
$\gamma_0 = 4 \omega^3 |\vec{d}|^2/3 \hbar c^3$,
and $\sigma_+$, $\sigma_-$ are the standard raising and lowering operators,
respectively given by
\begin{equation}
\sigma_+ = |1 \rangle \langle 0| =  {1 \over 2}
\left(\sigma_1 + i \sigma_2 \right);~~~
\sigma_- = |0 \rangle \langle 1| = {1 \over 2}
\left(\sigma_1 - i \sigma_2 \right).  \label{4c}
\end{equation}
Eq. (\ref{4b}) may be expressed in a manifestly Lindblad form as
\begin{equation}
\frac{d}{dt}\rho^s(t) = \sum_{j=1}^2\left(
2R_j\rho^s R^{\dag}_j - R_j^{\dag}R_j\rho^s - \rho^s R_j^{\dag}R_j\right),
\end{equation}
where $R_1 = (\gamma_0(N_{\rm th}+1)/2)^{1/2}R$,
$R_2 = (\gamma_0N_{\rm th}/2)^{1/2}R^{\dag}$ and 
$R = \sigma_-\cosh(r) + e^{i\Phi}\sigma_+\sinh(r)$. This observation guarantees
that the evolution of the density operator can be given a Kraus or
operator-sum representation \cite{nc00}, a point we return to later below.
If $T=0$, then $R_2$ vanishes, and a single Lindblad
operator suffices to describe Eq. (\ref{4b}).

In the above equation we use the nomenclature $|1 \rangle$ for the upper state
and $|0 \rangle$ for the lower state and $\sigma_1, \sigma_2, \sigma_3$
are the standard Pauli matrices.
In Eq. (\ref{4b}) 
\begin{eqnarray}
N &=& N_{\rm th}(\cosh^2(r) + \sinh^2(r)) + \sinh^2(r), \nonumber \\
M &=& -{1 \over 2} \sinh(2r) e^{i\Phi} (2 N_{\rm th} + 1), \nonumber \\
N_{\rm th} &=& {1 \over e^{{\hbar \omega \over k_B T}} - 1}. \label{4d}
\end{eqnarray}
Here $N_{\rm th}$ is the Planck distribution giving the number of thermal
photons at the frequency $\omega$ and $r$, $\Phi$ are squeezing parameters.
The analogous case of a thermal bath without squeezing can be obtained
from the above expressions by setting these squeezing parameters to zero.
We solve the Eq. (\ref{4b}) using the Bloch vector formalism 
to obtain the reduced density matrix of the system 
in the Schr\"{o}dinger picture as \cite{srigp} 
\begin{equation}
\label{eq:bmrhos}
\rho^s (t) = 
\left( \begin{array}{cc} 
{1 \over 2} (1 + A) & B e^{-i \omega t} 
\\ B^* e^{i \omega t} & 
{1 \over 2} (1 - A)
\end{array} \right), 
\end{equation}
where, 
\begin{eqnarray}
A \equiv \langle\sigma_3(t)\rangle
= e^{-\gamma_0 (2N + 1)t} \langle 
\sigma_3 (0) \rangle - \nonumber \\
{1 \over (2N + 1)} \left(1 - e^{-\gamma_0 (2N + 1)t} 
\right), \label{4m} 
\end{eqnarray}
\begin{eqnarray}
B &=& \left[1 + {1 \over 2} \left(e^{\gamma_0 a t}
- 1\right) \right] e^{-{\gamma_0 \over 2}(2N + 1 + a)t}
\langle \sigma_- (0) \rangle \nonumber \\
&+& \sinh({\gamma_0 a t \over 2}) e^{i \Phi 
- {\gamma_0 \over 2}(2N + 1)t} \langle \sigma_+ (0) \rangle. \label{4n}
\end{eqnarray}
Here $a=\sinh(2r)(2N_{\rm th}+1)$.
Making use of Eq. (\ref{4c}), Eq. (\ref{4n}) can be written as
$B = R e^{-i \chi}$. The explicit expressions for $R$ and $\chi$
may be found in Ref. \cite{srigp}.
For the determination of GP we need the eigenvalues and eigenvectors of 
the Eq. (\ref{eq:bmrhos}). The
eigenvalues are
\begin{equation}
\lambda_{\pm} (t) = {1 \over 2} \left( 1 + \epsilon_{\pm} \right), 
\label{4q4r}
\end{equation}
where $\epsilon_{\pm} = \pm \sqrt{A^2 + 4 R^2}$. 
As can be seen from the above expressions, at $t = 0$, $\lambda_+ (0) = 1$
and $\lambda_- (0) = 0$, hence for the purpose of GP we need only
the eigenvalue $\lambda_+ (t)$, and its corresponding normalized eigenvector
is given as
\begin{equation}
|\Psi_+ (t) \rangle = \sin\left({\theta_t \over 2}\right)
|1\rangle + e^{i (\chi (t) + \omega t)}\cos\left({\theta_t \over 2}\right)
|0\rangle, \label{4s}
\end{equation}
where 
$\sin\left(\theta_t/2\right) 
= {2 R \over \sqrt{4 R^2 + (\epsilon_+ - A)^2}}
= \sqrt{\frac{\epsilon_+ + A}{2\epsilon_+}}$.
It can be seen that for $t = 0$, $\chi(0)=\phi_0$,
$\sin\left({\theta_t \over 2}\right) =
\sqrt{{1 + \langle \sigma_3 (0) \rangle \over 2}} \equiv 
\cos\left({\theta_0 \over 2}\right)$ and $\cos\left({\theta_t \over 2}\right) =
\sqrt{{1 - \langle \sigma_3 (0) \rangle \over 2}} \equiv 
\sin\left({\theta_0 \over 2}\right)$, 
as expected. Now we make use of Eqs. (\ref{4q4r}),
(\ref{4s}) in Eq. (\ref{3g}) to obtain GP as 
\begin{eqnarray}
\Phi_{\rm GP} &=& \arg\Big[\{{1 \over 2} \left( 1 + 
\sqrt{A^2 (\tau) + 4 R^2(\tau)}\right)\}^{1 \over 2} \nonumber\\
&\times& \left\{\cos\left({\theta_0 \over 2}\right)
\sin\left({\theta_{\tau} \over 2}\right) \right. \nonumber \\
&+ & \left. e^{i (\chi(\tau) - \chi(0) + \omega \tau)}
\sin\left({\theta_0 \over 2}\right)
\cos\left({\theta_{\tau} \over 2}\right)\right\} \nonumber \\ 
&\times& e^{-i \int_0^{\tau} dt (\dot{\chi}(t) + \omega) 
\cos^2({\theta_t \over 2})}\Big]. \label{4v}
\end{eqnarray}
It can be easily seen from the Eq. (\ref{4v}) that if we set the influence
of the environment, encapsulated here by the terms $\gamma_0 , a$ and $\Phi$, 
to zero, we obtain for $\tau = {2 \pi \over \omega}$, $\Phi_{GP} =
-\pi(1- \cos(\theta_0))$, as expected,
which is the standard result 
for the unitary evolution of an intial pure state \cite{sjo00,sgh03}.
Thus we see that though the Eqs. (\ref{3m}), (\ref{4v})
represent the GP of a two-level system interacting with different kinds
of $S$-$R$ interactions, 
when the environmental effects are set to zero they yield
identical results. This is a nice consistency check for these expressions. 

\begin{figure}
\resizebox{0.95\columnwidth}{!}{\includegraphics{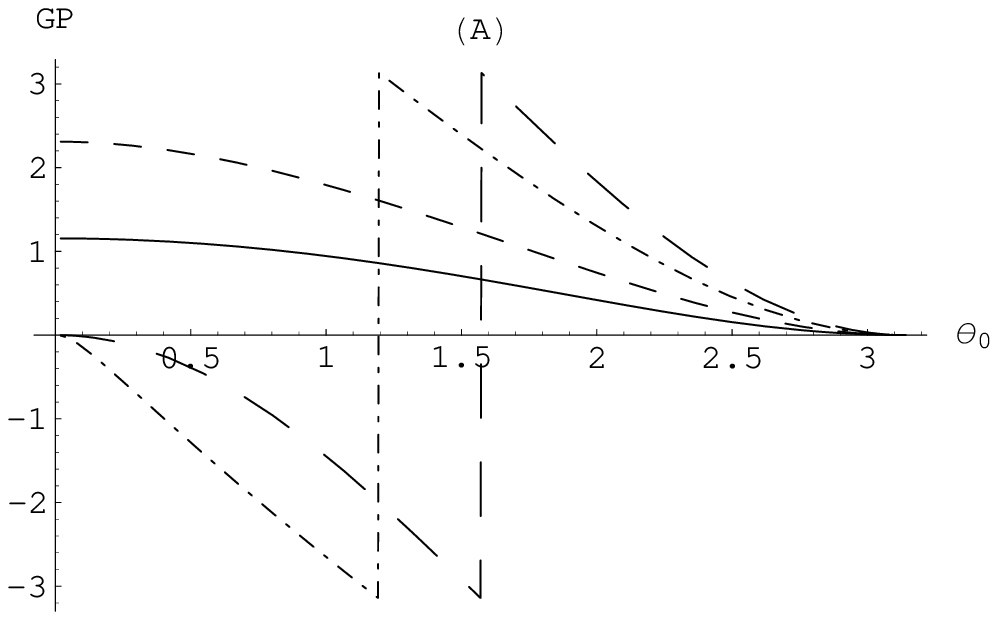}}
\hfill
\resizebox{0.95\columnwidth}{!}{\includegraphics{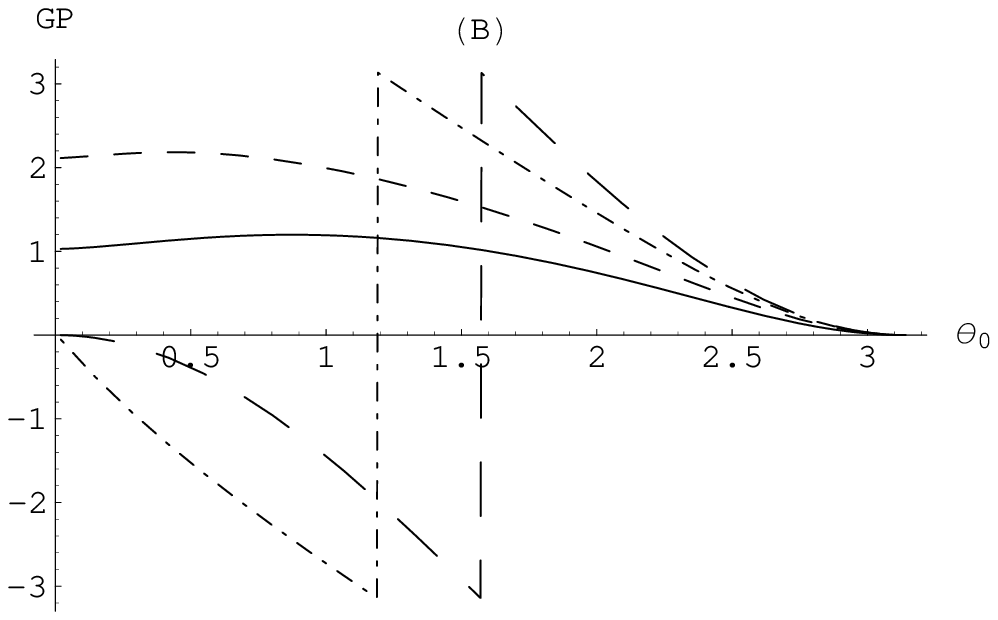}}
\caption{GP as a function of $\theta_0$ (in radians)
for different values of $\gamma_0$ and squeezing
in the Born-Markov approximation (Eq. (\ref{4v})). The discontinuity in
GP after $\pi$ is due to the convention that an angle in the third
quandrant is treated as negative. 
(A) $T=0$. The large-dashed curve
is the unitary case ($\gamma_0=0$). The dot-dashed 
(small-dashed) curve
represents $\gamma_0=0.1$ ($\gamma_0=0.3$). The solid curve 
represents $\gamma_0=0.6$. The stationary state, for
which GP vanishes, corresponds to $\theta_0 = \pi$
(i.e., $|0\rangle$), to which all states in the Bloch sphere
are asymptotically driven. Thus, 
a qubit started in this state
remains stationary and acquires no GP.
(B) Same as Figure (A), except that squeezing $r=0.4$, $\Phi=\pi/4$. }
\label{fig:bmTsq0}
\end{figure}
As expected, increasing the temperature,
$S{-}R$ coupling strength or squeezing induces a departure of GP from
unitary behavior. However the interpretation is less straightforward
than in the QND case. 
Further, introduction of squeezing complicates this pattern by
disrupting the monotonicity of the GP plots, as evident from the `humps'
seen for example in the Fig. \ref{fig:bmTsq0}(B), in comparison 
with those in Fig. \ref{fig:bmTsq0}(A). 

In all cases, we find that GP vanishes at $\theta_0=\pi$, i.e., for
a system that starts in the south pole of the Bloch sphere. On the
other hand, for sufficiently small $\gamma_0$, we find from 
Figs. \ref{fig:bmTsq0}(A) and \ref{fig:bmTsq0}(B) that GP may vanish
also in the case $\theta_0=0$.
These observations may be interpreted in the Bloch
vector picture, and are discussed in
Section \ref{sec:bma}.

\begin{figure}
\resizebox{0.95\columnwidth}{!}{\includegraphics{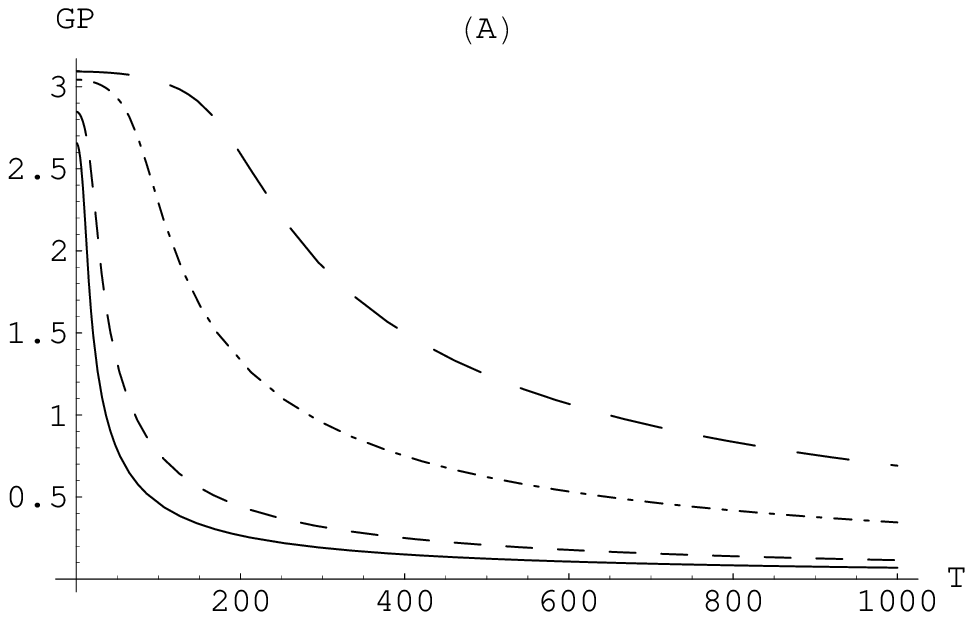}}
\hfill
\resizebox{0.95\columnwidth}{!}{\includegraphics{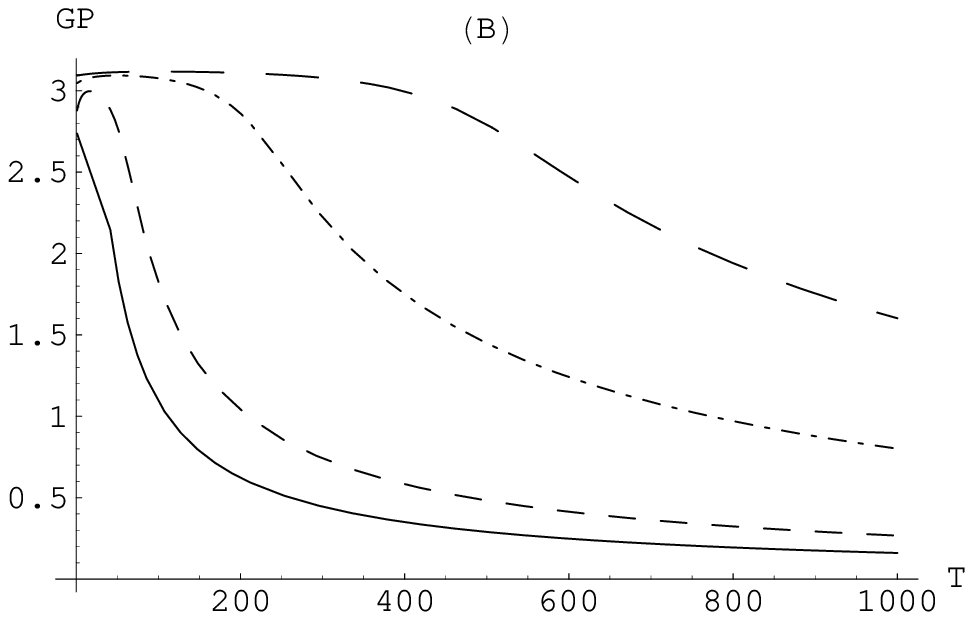}}
\caption{GP (in radians) vs temperature  ($T$, in units where
$\hbar \equiv k_B \equiv 1$) from Eq. (\ref{4v}).
Here $\omega=1.0$, $\theta_0=\pi/2$, the large-dashed, dot-dashed,
small-dashed and solid curves, represent, respectively,
$\gamma_0 = 0.005$, $0.01$, $0.03$ and $0.05$. (A)
squeezing is set to zero; (B) squeezing non-vanishing, with
$r=0.4$ and $\Phi=0$.}
\label{fig:zanHalfpi}
\end{figure}

\begin{figure}
\resizebox{0.95\columnwidth}{!}{\includegraphics{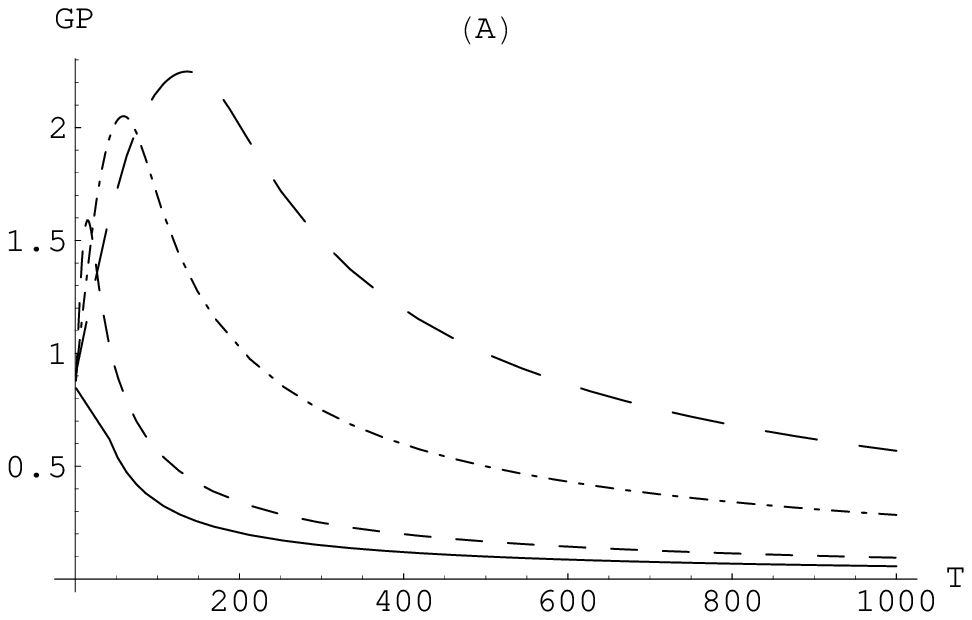}}
\hfill
\resizebox{0.95\columnwidth}{!}{\includegraphics{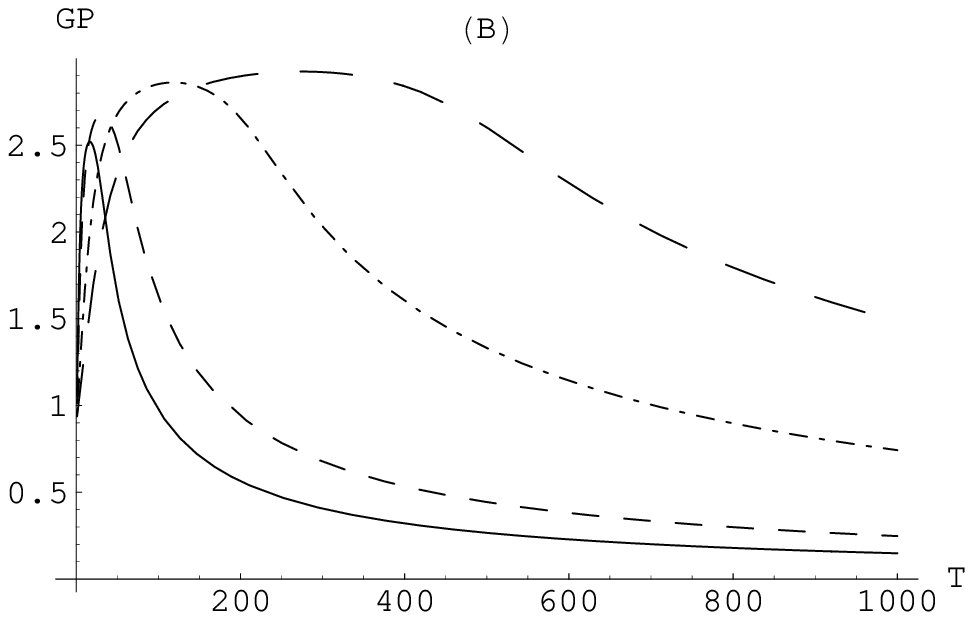}}
\caption{GP vs temperature ($T$, in units where
$\hbar \equiv k_B \equiv 1$) from Eq. (\ref{4v}).
Here $\omega=1.0$, $\theta_0=\pi/2 + \pi/4$. The curves
represent $\gamma_0 = 0.005$, $0.01$, $0.03$ and $0.05$ as in
Fig. \ref{fig:zanHalfpi}. (A)
squeezing is set to zero; (B) squeezing non-vanishing, with
$r=0.4$ and $\Phi=0$.}
\label{fig:zanpi2pi4}
\end{figure}

In contrast  to the situation  in a purely  dephasing system, GP  in a
dissipative  system  is  rather  complicated,  and  less  amenable  to
interpretation.  The  dependence of GP  on temperature is  depicted in
Figs.   \ref{fig:zanHalfpi}  and  \ref{fig:zanpi2pi4}.   The  expected
pattern  of GP falling  asymptotically with  temperature is  seen. Our
results  parallel those  obtained in  Refs. \cite{rz06,mar04}  for the
case   of   zero    squeezing   (Figs.    \ref{fig:zanHalfpi}(A)   and
\ref{fig:zanpi2pi4}(A)),  and extend them  to the  case of  a squeezed
thermal environment.  We note that  the effect of squeezing is to make
GP  vary more  slowly with  temperature,  by broadening  the peak  and
fattening  the tails  of the  plots.  This  counteractive  behavior of
squeezing on  the influence  of temperature  on GP for  the case  of a
dissipative system  is interesting,  and would be  of use  in practial
implementation  of   geometric  phase  gates.   This   effect  can  be
understood by visualizing the  effects of squeezing and temperature on
the Bloch sphere, a point we return to in Section \ref{sec:bma}.

\subsection{Evolution of GP in a squeezed generalized
amplitude damping channel  \label{sec:bma}}

While the  results derived  in this section  pertain to  a dissipative
$S$-$R$ interaction  in the Born-Markov  RWA, they are  quite general,
and are applicable to any open system effect that can be characterized
as  a squeezed  generalized amplitude  damping  channel \cite{srisub}.
Amplitude damping channels capture the idea of energy dissipation from
a system,  for example,  in the spontaneous  emission of a  photon, or
when a spin system at high temperature approaches equilibrium with its
environment. A  simple model  of an amplitude  damping channel  is the
scattering of a photon via a beam-splitter. One of the output modes is
the environment,  which is traced  out. The unitary  transformation at
the  beam-splitter is  given  by $B  =  \exp\left[\theta (a^{\dag}b  -
ab^{\dag})\right]$,  where  $a,  b$  and $a^{\dag},b^{\dag}$  are  the
annihilation and creation operators for photons in the two modes.  The
generalized  amplitude damping channel,  with $T\ge  0$ and  with zero
squeezing, extends the amplitude damping channel to finite temperature
\cite{nc00}. A  very general  CP map generated  by Eq.  (\ref{4b}) has
been recently obtained by us \cite{srisub}, and could be appropriately
called  the  squeezed  generalized  amplitude  damping  channel.  This
extends  the generalized  amplitude  damping channel  by allowing  for
finite  bath squeezing.  It  is characterized  by the  Kraus operators
\cite{srisub}
\begin{eqnarray}
E_0 &\equiv& \sqrt{p_1}\left[\begin{array}{ll} 
\sqrt{1-\alpha(t)} & 0 \\ 0 & 1
\end{array}\right],\nonumber \\
E_1 &\equiv& \sqrt{p_1}\left[\begin{array}{ll} 0 & 0 
\\ \sqrt{\alpha(t)} & 0 \end{array}\right], \nonumber \\
E_2 &\equiv& \sqrt{p_2}\left[\begin{array}{ll} 
\sqrt{1-\mu(t)} & 0 \\ 0 & \sqrt{1-\nu(t)}
\end{array}\right], \nonumber \\
E_3 &\equiv& \sqrt{p_2}\left[\begin{array}{ll} 0 & \sqrt{\nu(t)} 
\\ \sqrt{\mu(t)}e^{-i\Phi}  & 0
\end{array}\right]. 
\label{eq:gbmakraus}
\end{eqnarray}
With some  algebraic manipulation,  it can be  verified that  with the
identification
\begin{eqnarray}
\label{eq:munulfa}
\nu(t) &=& \frac{N}{p_2(2N+1)}(1-e^{-\gamma_0(2N+1)t}), \nonumber \\
\mu(t) &=& \frac{2N+1}{2p_2 N}
\frac{\sinh^2(\gamma_0at/2)}{\sinh(\gamma_0(2N+1)t/2)} 
\exp\left(-\frac{\gamma_0}{2}(2N+1)t\right),\nonumber \\
\alpha(t) &=& \frac{1}{p_1}\left(1 - p_2[\mu(t)+\nu(t)]
- e^{-\gamma_0(2N+1)t}\right),
\end{eqnarray}
where $N$ is as  in Eq. (\ref{4d}), the operators (\ref{eq:gbmakraus})
acting   on    the   state   (\ref{3e})    reproduce   the   evolution
(\ref{eq:bmrhos}), by means of the map Eq. (\ref{eq:kraus}), provided
$p_2 = 1-p_1$, satisfies
\begin{eqnarray}
p_2 &=& \frac{1}{(A+B-C-1)^2-4D} \\
& \times&  \left[A^2B + C^2 + A(B^2 - C - B(1+C)-D)\right. \nonumber \\
&-& \left. (1+B)D - C(B+D-1) 
\nonumber \right. \\
&\pm&   2\left(D(B-AB+(A-1)C+D)\right. \nonumber \\
&\times& \left.\left. (A-AB+(B-1)C+D)\right)^{1/2}\right],
\nonumber
\label{eq:p2}
\end{eqnarray}
where
\begin{eqnarray}
A &=& \frac{2N+1}{2N} \frac{\sinh^2(\gamma_0 at/2)}
{\sinh(\gamma_0(2N+1)t/2)}
\exp\left(-\gamma_0(2N+1)t/2\right), \nonumber \\
B &=& \frac{N}{2N+1}(1-\exp(-\gamma_0(2N+1)t)), \nonumber \\
C &= & A + B + \exp(-\gamma_0 (2N+1)t), \nonumber \\
D &=& \cosh^2(\gamma_0 at/2)\exp(-\gamma_0(2N+1)t).
\label{eq:auxip2}
\end{eqnarray}

As  the  interaction  in  the  Born-Markov  RWA  realizes  a  squeezed
generalized  amplitude  damping  channel  \cite{srisub},  the  various
qualitative features  of GP seen under a  dissipative interaction (for
example, the relatively complicated  dependence of GP on $\theta_0$, and
on evolution  time) carry over  to any squeezed  generalized amplitude
damping channel.  If squeezing parameter $r$ is set to zero, it can be
seen from above that Eq. (\ref{eq:gbmakraus}) reduces to a generalized
amplitude damping  channel, with $\nu(t) =  \alpha(t)$, $\mu(t)=0$ and
$p_1$ and $p_2$  being time-independent.  If further $T=0$,  it can be
seen from  above that  $p_2=0$, reducing Eq.   (\ref{eq:gbmakraus}) to
two Kraus operators, corresponding to an amplitude damping channel.

Refs. \cite{wang06} and \cite{wang07} consider GP evolving under 
an amplitude damping channel and a squeezed amplitude damping
channel, respectively. These are subsumed under the
squeezed generalized amplitude damping channel considered above.
This channel is contractive,
in that the system is seen to evolve
towards a fixed asymptotic point in the Bloch sphere, 
which in general is not a pure state,
but the mixture
\begin{equation}
\rho_{\rm asymp}
= \left(\begin{array}{ll} 1-q & 0 \\ 0 & q \end{array} \right),
\end{equation}
where $q = (N+1)/(2N+1)$.  If  $T=r=0$, then $q=1$, and the asymptotic
state is the pure state $|0\rangle$. Physically this can be understood
as a system going to its  ground state by equilibriating with a vacuum
bath, This can have a  practical application in quantum computation in
the form  of a quantum  deleter \cite{qdele}. At  $T=\infty$, $p=1/2$,
and the system  tends to a maximally mixed  state, thereby realizing a
fully depolarizing channel \cite{nc00}.
\begin{figure}
\resizebox{0.75\columnwidth}{!}{\includegraphics{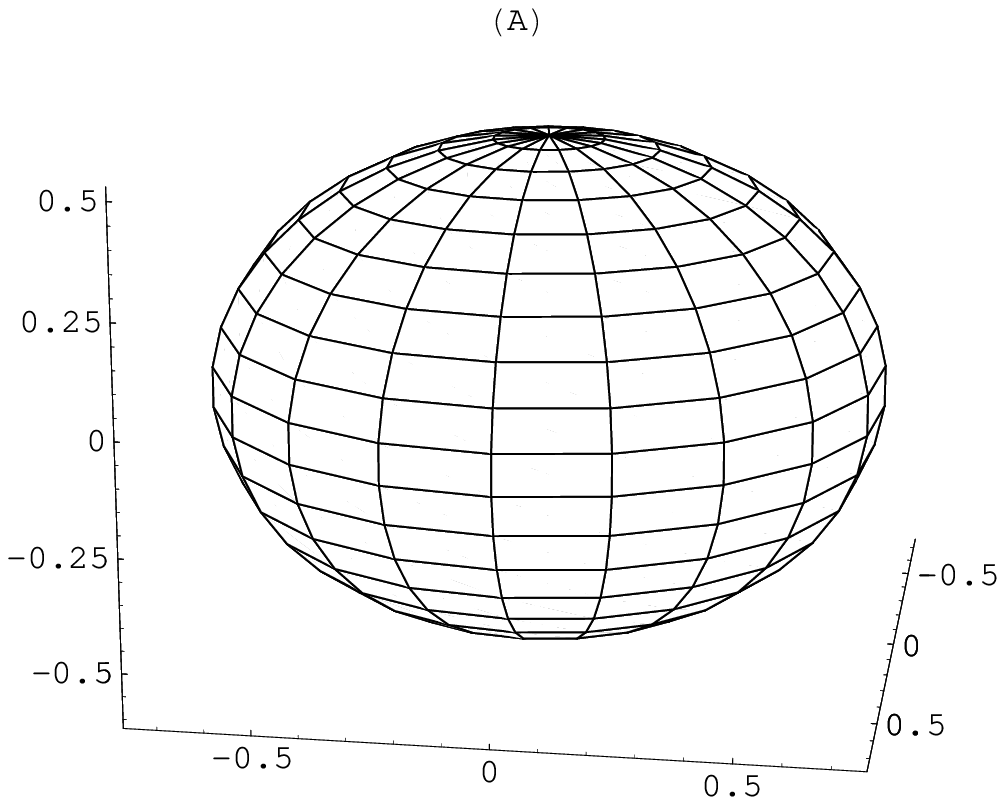}}
\hfill
\resizebox{0.75\columnwidth}{!}{\includegraphics{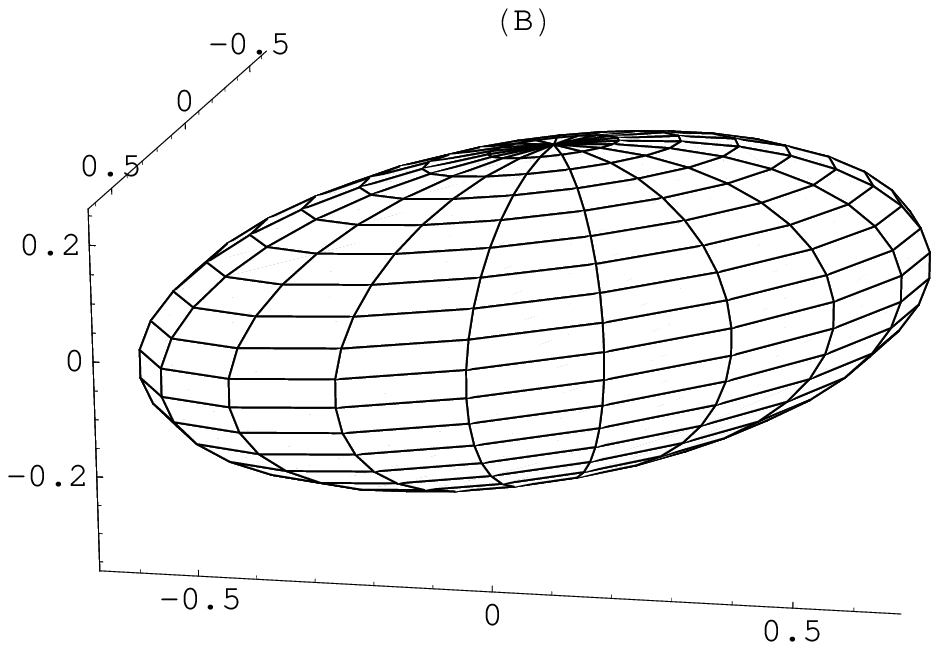}}
\caption{Shrinking of  the full Bloch  sphere into an  oblate spheroid
under evolution given by a Born-Markov type of dissipative interaction
with $\gamma_0=0.6$ and temperature $T=5.0$.  In (B), the $x$-$y$ axes
are interchanged for  convenience.  (A) $r = \Phi  = 0$, $t=0.15$; (B)
$r=0.4,  \Phi=1.5$, $t=0.15$.   Finite $\Phi$  is responsible  for the
tilt. }
\label{fig:bm_skrink}
\end{figure}

As  in the  case of  the QND  interaction, abstracting  the  effect of
dissipative  interaction into  the Kraus  representation allows  us to
subsume all the details of the system into a limited number of channel
parameters $p_1(t)$, $\Phi$,  $\alpha(t)$, $\mu(t)$ and $\nu(t)$.  Any
other  dissipative system  that can  be described  by  a Lindblad-type
master  equation  Eq.  (\ref{4b})  will  show  a  similar  pattern  in
behavior.

To develop  physical insight  into the solution,  we transform  to the
interaction picture, and for  simplicity, set the squeezing parameters
to  zero.  Then,  the  action of  the operators  (\ref{eq:gbmakraus}),
[which now represents a generalized amplitude channel] on an arbitrary
qubit state is given in the Bloch vector representation by
\begin{eqnarray}
\label{eq:gbma}
\langle\vec{\sigma}(t)\rangle &=& 
(\langle\sigma_1(0)\rangle\sqrt{1-\lambda(t)},
\langle\sigma_2(0)\rangle\sqrt{1-\lambda(t)},\nonumber \\ 
&& \lambda(t)(1-2p) + \langle\sigma_3(0)\rangle(1-\lambda(t))),
\end{eqnarray}
where $p  = (N_{\rm  th}+1)/(2N_{\rm th}+1)$ and  $\lambda(t=\infty) =
1$.   Thus, the Bloch  sphere contracts  towards the  asymptotic mixed
state $(0,0,1-2p)$ (Fig.  \ref{fig:bm_skrink}(A)), characteristic of a
generalized  amplitude  damping  channel,   with  $T  \ge  0$  and  no
squeezing.  If  $T=0$  case,  then  $p=1$, and  the  asymptotic  state
$(0,0,-1)$ is pure.

The Bloch vector picture allows us to interpret the results of Section
\ref{sec:gpbma}.  Eqs. (\ref{eq:bmrhos}),  show that the Bloch vector
for the  states corresponding to $\theta_0  = 0, \pi$  move only along
the $z$-axis of the Bloch sphere  for zero as well as finite $T$.  For
the  case  $\theta_0=\pi$  and  zero  $T$, the  Bloch  vector  remains
stationary at  $(0,0,-1)$, and  hence GP vanishes.  In the  finite $T$
case,  GP  still vanishes,  because  the  Bloch  vector has  the  form
$(0,0,-L(t))$,  where the Bloch  vector length  $L(t)$ shrinks  from 1
towards an  interaction-dependent asymptotic value, which  is zero for
infinite  temperature  or finite  otherwise.  Since  the Bloch  vector
shrinks strictly along  its length, and thus subtends  no finite angle
at  the   center  of  the  sphere,   we  find  that   GP  vanishes  at
$\theta_0=\pi$, as expected (cf. Figs. \ref{fig:bmTsq0}).

On  the other  hand, even  though the  Bloch vector  shrinks similarly
along  its  length  in the  case  $\theta_0=0$,  we  find that  GP  is
non-vanishing  in certain  cases, in  fact, in  precisely  those cases
where the  tip of  the Bloch  vector crosses the  center of  the Bloch
sphere moving  along the $\sigma_3$-axis. That is,  they correspond to
the  situation  where  $\langle\sigma_3(t)\rangle$ changes  sign  from
positive to negative  during the period of one  cycle. In these cases,
the dependence  of GP on  the Bloch vector  is too involved for  us to
interpret in terms of $L$ and the angle subtended by the Bloch vector,
for  some  qualitative  insight.  Nevertheless  this  feature  may  be
formally understood as follows. It can be observed from Eq. (\ref{4m})
that  for sufficiently  large  $\gamma_0$, $\langle\sigma_3(t)\rangle$
changes sign  at $t_1 \equiv  \log(2[N+1])/(\gamma_0[2N+1])$. Further,
we note that  $R$ vanishes for $\theta_0 = 0$ (as  well as $\theta_0 =
\pi$).

It is convenient to recast Eq. (\ref{4v}) in the expanded form
\begin{eqnarray}
\label{eq:newGP}
\Phi_{\rm GP} &=& \tan^{-1}\left[
\left(\sin(\chi(\tau)-\chi(0)+2\pi)
\sin(\theta_0/2)\cos(\theta_{\tau}/2)\right) \right. \nonumber \\
&\div& \{\cos(\chi(\tau)-\chi(0)+2\pi)
\sin(\theta_0/2)\cos(\theta_{\tau}/2) \nonumber \\
&+& \left. \cos(\theta_0/2)\sin(\theta_{\tau}/2)\} \right] 
\nonumber \\
&-& \int_0^{\tau} dt (\dot{\chi}(t) + \omega) 
\cos^2\left(\frac{\theta_t}{2}\right). 
\end{eqnarray} 
It can be seen that
for the case $\theta_0=\pi$, $\cos(\theta_t/2)=1$ and, in particular,
$\cos(\theta_{\tau}/2)=1$. Substituting these values in Eq. 
(\ref{eq:newGP}), it is seen that GP vanishes because the two terms
in the RHS of Eq. (\ref{eq:newGP}) cancel each other. Next consider the
case where $\theta_0=0$ but where $\gamma_0$ is sufficiently weak
that $\tau \le t_1$, i.e., $\langle \sigma_3(t)\rangle$ 
does not change sign during one
cycle. In this case, from above it is seen
that $\cos(\theta_{t}/2)=0$, and, in particular, $\cos(\theta_{\tau}/2)=0$,
and thus the terms in the RHS
of Eq. (\ref{eq:newGP}) vanish identically. But in the case of
$\theta_0=0$ where $\tau > t_1$ ($\gamma_0$ being relatively
stronger), $\cos(\theta_{t}/2)=0$
initially in the time interval $[0,t_1]$, and then switches 
to 1 in the interval $(t_1,\tau]$. In particular,
$\cos(\theta_{\tau}/2)=1$. Observe that if $\cos(\theta_{t}/2)=1$
throughout the interval $[0,\tau]$, the two terms in the RHS cancel
each other. It follows that GP is non-vanishing because of an
excess contributed by the first term, in the interval $[0,t_1]$.

Contraction produced by an increase in temperature tends to be less 
pronounced in the presence (than in the absence) of squeezing
(Figs. \ref{fig:bm_skrink}). This is reflected in the 
slower variation of GP with respect to temperature, seen in 
Figs. \ref{fig:zanHalfpi}(B) and \ref{fig:zanpi2pi4}(B) in relation to
Figs. \ref{fig:zanHalfpi}(A) and \ref{fig:zanpi2pi4}(A), respectively.
As observed in Figs. \ref{fig:zanHalfpi} and \ref{fig:zanpi2pi4},
GP falls as a function of $T$, for sufficiently large $T$. This
may quite generally be attributed to the reduction in $L$ and $\Omega$ caused
by the contraction of Bloch vector 
as a result of interaction with the environment.
The tilt of the contracted Bloch sphere in Fig. \ref{fig:bm_skrink}(B)
is due to finite $\Phi$. 

\section{Conclusions\label{sec:discu}}
We have studied the combined influence of squeezing and temperature on
the GP for  a qubit interacting with a bath  both in a non-dissipative
as well as in a dissipative  manner. In the former case, squeezing has
a similar  debilitating effect as  temperature on GP. In  contrast, in
the latter case, squeezing can counteract the effect of temperature in
some regimes.  This makes  squeezing potentially helpful for geometric
quantum   information  processing   and  geometric   computation.   In
particular,  in  the  context  of using  engineered  (e.g.,  squeezed)
reservoirs  to  generate  GP  \cite{cor06},  it would  be  helpful  to
consider  the  effect  of  squeezing  together  with  thermal  effects
\cite{rz06,lom06}.

In the non-dissipative (QND) case, we analyzed a number of open system
models using two types of bath: the usual one of harmonic oscillators,
and that of two-level systems.  It was shown that for the case of weak
$S{-}R$  coupling, the  two kinds  of baths  can be  mapped  onto each
other.   GP was  studied  as a  function  of the  initial polar  angle
$\theta_0$  of the  Bloch sphere,  temperature and  squeezing (arising
from the  squeezed thermal bath).  In  the QND case, it  was seen that
increasing  $\gamma_0$,  temperature or  squeezing  tends  to cause  a
similar departure from unitary behavior by suppressing GP.

However, in the  dissipative case (with the environment  modelled as a
squeezed thermal bath in the  weak Born-Markov RWA), we found that the
dependence  of GP  on $\theta_0$,  temperature and  squeezing  shows a
greater complexity.  Here, an  interesting feature due to squeezing is
that  it  can  disrupt,  over  an interval,  the  otherwise  monotonic
behavior of GP  as a function of $\theta_0$ (the  humps seen in Figure
\ref{fig:bmTsq0}(B)).  More pronouncedly,  the counteractive effect of
squeezing on  temperature is  brought out by  a comparison  of Figures
\ref{fig:zanHalfpi}(A)      with      \ref{fig:zanHalfpi}(B),      and
\ref{fig:zanpi2pi4}(A) with  \ref{fig:zanpi2pi4}(B).  Also, its effect
on the Bloch sphere is to shrink it to an oblate spheroid, in contrast
to a QND interaction, which produces a prolate spheroid. Thus, an
interesting feature that emerges from  our work is the contrast in the
interplay between squeezing and thermal effects in non-dissipative and
dissipative  interactions.   By
interpreting the open  quantum effects as noisy channels,  we make the
connection  between  geometric   phase  and  quantum  noise  processes
familiar from quantum information theory.

An added feature of our work  is that we make a connection between the
studied  open  system models  and  the  phase  damping and  the  newly
introduced   squeezed  generalized  amplitude   damping  \cite{srisub}
channels,  noise   processes  which  are  important   from  a  quantum
information  theory perspective.   In particular,  we give  a detailed
microscopic basis for  these noisy channels.  This allows  us to study
the effects of the formal noise processes on GP.

\end{document}